# Modeling the Input History of Programs for Improved Instruction-Memory Performance

Carlos A. G. Assis
Edil S. T. Fernandes
Valmir C. Barbosa*

Universidade Federal do Rio de Janeiro
Programa de Engenharia de Sistemas e Computação, COPPE
Caixa Postal 68511
21941-972 Rio de Janeiro - RJ, Brazil

November 23, 2004

## Abstract

When a program is loaded into memory for execution, the relative position of its basic blocks is crucial, since loading basic blocks that are unlikely to be executed first places them high in the instruction-memory hierarchy only to be dislodged as the execution goes on. In this paper we study the use of Bayesian networks as models of the input history of a program. The main point is the creation of a probabilistic model that persists as the program is run on different inputs and at each new input refines its own parameters in order to reflect the program's input history more accurately. As the model is thus tuned, it causes basic blocks to be reordered so that, upon arrival of the next input for execution, loading the basic blocks into memory automatically takes into account the input history of the program. We report on extensive experiments, whose results demonstrate the efficacy of the overall approach in progressively lowering the execution times of a program on identical inputs placed randomly in a sequence of varied inputs. We provide results on selected SPEC CINT2000 programs and also evaluate our approach as compared to the gcc level-3 optimization and to Pettis-Hansen reordering.

*Corresponding author (`valmir@cos.ufrj.br`).

# 1 Introduction

It is a well-known fact that only a small fraction of a program's instructions is responsible for most of its running time. Coupled with the growing gap that exists between memory and processor performance [16], this has over the years led to the search for code-layout techniques for optimizing the use of the memory system. The essential guiding principle in this search is that the first of a program's basic blocks to be loaded into memory should be precisely those that are most likely to be executed.

The earliest efforts related to optimizing code layout concentrated on virtual-memory systems [4, 6, 7] and aimed at producing code layouts that could reduce the number of page faults at runtime. The advent of TLB's and the introduction of several cache levels in recent processors have both shifted the context considerably and added new momentum to the search for efficient techniques. Naturally, the focus of this search is invariably placed on the investigation of heuristic techniques, since the optimality of a code layout cannot in general be decided [1].

Notable contributions within this more recent context include some that target the reduction of the instruction-cache miss rate [9, 8, 12], or the reduction of cache pollution and bus traffic [5], or yet the reduction of the program's running time [11, 2, 14]. Most of these contributions involve instrumenting the program for trace recording and the eventual construction of a profile on which the code-layout optimization technique operates. Some others, however, concentrate solely on the development of new compile-time techniques or combine profiling with compilation strategies.

Our interest in this paper is to investigate the construction of a dynamic probabilistic model of the inputs to a program. Specifically, as the program is run on an assorted sequence of inputs, we describe how a probabilistic model can be dynamically updated so that, at all times, it reflects the input history of the program and can as such be used to update the program's code layout for subsequent use. Like in so many of the techniques mentioned above, this model is built on the profile of the program's execution on each input. Unlike those techniques, however, the one we introduce is based on a model that persists from one execution of the program to the next while refining itself as information on each new execution comes in. What we ultimately seek is the improvement of running times with as much generality as possible (this includes, for example, independence from specific cache sizes, once again in contrast to some of the related approaches).

The following is a high-level description of our overall strategy. Let $P$ be the binary code of some program for the architecture at hand, and $I, I'$ two independent inputs to $P$. Suppose we have a means of instrumenting $P$ so that running it on input $I$ yields an abstract model of this particular execution which can be used to estimate the set of basic blocks of $P$ that is most likely to occur in future executions. If this is the case, then we can reorder the basic blocks of $P$ in such a way that, when input $I'$ comes along for execution, the basic blocks to occupy the highest levels in the instruction-memory hierarchy are none other than some of those that were previously estimated to be the most likely to occur. That estimate, of course, was based solely on input $I$, so it may work rather poorly on the new input $I'$. However, the execution of $P$ on $I'$ can itself be instrumented and the model resulting from this second execution can be combined with the previous one in the hope of a more general, less execution-dependent model to be used in a subsequent run of $P$.

Our central premise in this paper is that such modeling of executions of $P$ can indeed be achieved and used successfully toward progressively more efficient runs of $P$ as it is applied to a stream of inputs. The model that we build of a particular execution of $P$ is based on recording a trace of the execution as it goes through the basic blocks of $P$ and then using the data in the trace to construct a Bayesian network [10, 3, 13]. Combining this Bayesian network with another that records a history of all previous executions of $P$, and then solving the resulting Bayesian network for the most likely combination of basic blocks, is what gives us the prediction capability that allows for progressively more efficient executions. Sections 2 and 3 contain, respectively, the details of our model-building and -updating methodology, and a summary of our overall strategy. Section 4 contains the results of extensive experimentation on the SPEC CINT2000 suite [15].

One immediate difficulty with this approach is of course that it may take considerable effort for the refined predictive model to be obtained from an execution: not only does instrumenting $P$ slows it down significantly, but also setting up the Bayesian network and solving it may be quite time-consuming. The entire strategy would then seem to be wholly inappropriate for a real-world environment, since any gain that might eventually be accrued would be totally overshadowed by the cost to obtain it. But we envisage a different dynamics for the successful application of our approach, one that only applies it to a substream of the stream of inputs to $P$, in such a way that rearranged versions of $P$ only become available every so often, as opposed to becoming available right after every new input is processed. We provide further considerations on this in Section 5 along with conclusions.

## 2 The model

We consider a sequence $I_1, I_2, \ldots$ of inputs to $P$, along with a corresponding sequence of directed graphs $G_1, G_2, \ldots$, where for $i \geq 1$ graph $G_i$ represents the recorded trace of executing $P$ on input $I_i$. Each node of $G_i$ is a basic block of $P$ that is reached during that execution. A directed edge exists in $G_i$ from basic block $a$ to basic block $b$, denoted by $(a \to b)$, if during the execution $b$ follows $a$ immediately at least once. In this case, the trace is complemented by a positive count, denoted by $f_{ab}$, indicating the number of times this happens. Notice that the node sets of $G_1, G_2, \ldots$ are not necessarily the same, even though they are all subsets of the set of $P$'s basic blocks.

Transforming each of these edge-labeled directed graphs into a Bayesian network is one of the crucial steps of modeling the executions of $P$ that they stand for. Before describing how the transformation is achieved, we pause briefly for a discussion of the basic principles of Bayesian networks.

## 2.1 Bayesian-network basics

A Bayesian network is a node-labeled acyclic directed graph whose nodes are random variables and whose directed edges represent the existence of direct causal influences. In other words, if $X$ and $Y$ are nodes, then the existence of the directed edge $(X \to Y)$ indicates that the value of $X$ influences the value of $Y$ directly. We use $\Pi_X$ to denote the set of variables from which edges exist directed toward $X$ (the so-called parents of $X$ in the Bayesian network). If we let $\boldsymbol{\pi}_X$ denote a joint value assignment to the variables in $\Pi_X$, then for each possible $\boldsymbol{\pi}_X$ the label that goes with node $X$ to complete the definition of the Bayesian network includes the conditional probability $p(x \mid \boldsymbol{\pi}_X)$ that $X$ has value $x$ given the values of $X$'s parents appearing in $\boldsymbol{\pi}_X$. In the case of $0,1$-variables, we need $2^{|\Pi_X|}$ such probabilities (which may be problematic, depending on the size of $\Pi_X$); if $\Pi_X = \emptyset$, then the single necessary probability is known as the prior probability of $X$.

One facilitating assumption that is always made in the study of Bayesian networks is that conditioning the value of $X$ on the values of the variables in $\Pi_X$ is the same as conditioning on the values of all the variables that cannot be reached from $X$ along directed paths. Given this assumption, and letting $\mathbf{x}$ denote a joint value assignment to all the variables in the Bayesian network, it is simple to see that

$$p(\mathbf{x}) = \prod_{X \in \mathbf{X}} p(x \mid \boldsymbol{\pi}_X), \tag{1}$$

where $\mathbf{X}$ is the set of all variables (the node set of the Bayesian network), $x$ is the value assigned to $X$ in $\mathbf{x}$, and $\boldsymbol{\pi}_X$ comprises the values assigned in $\mathbf{x}$ to the variables in $\Pi_X$.

In the context of this paper, the key problem to be solved once the Bayesian network has been set up is the following. Let $\mathbf{E} \subset \mathbf{X}$ comprise variables whose values are no longer uncertain but known with certainty. These are the so-called evidences and the problem asks for the joint value assignment to the variables in $\mathbf{X} \setminus \mathbf{E}$ that maximizes $p(\mathbf{x} \setminus \mathbf{e} \mid \mathbf{e})$, where $\mathbf{x} \setminus \mathbf{e}$ denotes one such joint value assignment and $\mathbf{e}$ the evidences' values.

This and other similar problems are in general computationally intractable, in the sense of NP-hardness, even though $p(\mathbf{x} \setminus \mathbf{e} \mid \mathbf{e})$ can be derived from (1) rather straightforwardly. This inherent difficulty stems essentially from the existence of multiple paths joining two nodes in the undirected graph that underlies the Bayesian network, and also from the absence of a constant bound on the sizes of the $\Pi_X$ sets.

There are several approximation schemes that can be used. The one we use in this paper is based on recognizing first that $p(\mathbf{x} \setminus \mathbf{e} \mid \mathbf{e})$ is proportional, by a normalizing constant, to the $p(\mathbf{x})$ of (1), and further that maximizing $p(\mathbf{x})$ over the possibilities for $\mathbf{x} \setminus \mathbf{e}$ is equivalent to minimizing the function

$$-\sum_{X \in \mathbf{X}} \ln p(x \mid \boldsymbol{\pi}_X) \tag{2}$$

over the same possibilities when the distribution in (1) is everywhere strictly positive.

This minimization, in turn, can be achieved by a variation of stochastic simulation that employs simulated annealing in an attempt at near-optimality. If $T$ is the temperature-like parameter of simulated annealing, then whenever during the process variable $X$ is to be updated, it is assigned value $x$ with probability proportional (by a normalizing constant) to

$$\prod_{Y \in \mathbf{N}_X} p(y \mid \boldsymbol{\pi}_Y)^{1/T}, \tag{3}$$

given $x$ as the value of $X$ and the current joint value assignment to some of the other variables in $\mathbf{X}$. In (3), $\mathbf{N}_X$ comprises $X$ itself and its so-called children (variables $Y$ such that $(X \to Y)$ is an edge); indirectly, then, the probability depends only on $x$, on the current values of $X$'s parents and children, and also on the current values of its children's other parents. These are all well-known results and for details we refer the reader to the pertinent literature [3]. Our use of the technique in this paper is concentrated in Section 4, where the necessary details are filled in.

As one last remark, notice that approximation schemes like the one we just outlined do nothing to handle the potentially problematic sizes of the $\Pi_X$ sets as far as storing labels that depend on such sets is concerned. The issue is crucial, though, and we return to it shortly.

## 2.2 The execution model

For $i \geq 1$, we model the execution of program $P$ on input $I_i$ by a Bayesian network denoted by $B_i$. Constructing $B_i$ involves transforming the edge-labeled directed graph $G_i$ into the node-labeled, acyclic directed graph $B_i$. We describe this transformation as a sequence of two steps. The first step transforms $G_i$ into an acyclic directed graph $G'_i$ that already has the desired structure of $B_i$ but still carries integer labels on its edges. The second steps completes the transformation into $B_i$ by computing its node labels (sets of conditional probabilities) from the edge labels of $G'_i$.

The node set of $G'_i$ contains one random variable for each of the nodes of $G_i$, that is, for each of the basic blocks of $P$ that is executed when $P$ is run on input $I_i$. A variable may only have value 0 or 1, representing respectively the event that the corresponding basic block is not or is executed in a run of $P$. Notice that this does in no way contradict the fact that, by definition, all basic blocks in $G_i$ are executed in the run of $P$ to which it corresponds. Building the probabilistic model $B_i$ is simply an intermediate step that seeks to later integrate the contribution of this particular run into a model of the input history of $P$. We let $X_a$ be the variable corresponding to basic block $a$.

The node sets of $G_i$ and $G'_i$ are thus so far in one-to-one correspondence with each other. Their edge sets, on the other hand, cannot in general have the same property, since $G'_i$ (having the same structure of $B_i$) must be acyclic, while $G_i$ is in general not so. The source of directed cycles in $G_i$ is of course the presence of basic blocks whose last instruction is a branch instruction to implement a loop in $P$. The edges of $G_i$ that correspond to such branches are precisely the ones that get eliminated in order to generate $G'_i$, which is then acyclic.

The process of eliminating branch edges is very simple. First we identify the only possible source node in $G_i$, i.e., the single node that has no edges incoming to it. This node represents the first basic block that is executed when $P$ is run on $I_i$; since $P$ is fixed, such a node is the same for all values of $i$. Once the source node is identified, a depth-first search is conducted starting at it and all the back edges it produces are eliminated. Note that, even though it can be argued that these back edges correspond precisely to the branch edges that implement loops in $P$, what matters is simply that the resulting directed graph is acyclic.

But let us examine the process of eliminating a back edge from $G_i$ more closely. By definition of the edge labels of $G_i$, summing up the labels on the edges incoming to any of its nodes (with the exception of its single source and its sinks—nodes with no outgoing edges) must yield the same value as summing up the labels on that node's outgoing edges. If $(a \rightarrow b)$ is a back edge, severing it disrupts this balance so that the resulting graph no longer conveys the same information as $G_i$ regarding the relative frequencies with which basic blocks are executed. What we do to solve this is to create two additional nodes (one source and one sink), called $a'$ and $b'$, and two additional edges, $(a' \rightarrow b)$ and $(a \rightarrow b')$, each receiving the same label, $f_{ab}$, of the edge being severed. Clearly, the desired balance is thus maintained. The resulting node and edge sets are shared by both $G'_i$ and $B_i$.

Let us now consider the second step in turning $G_i$ into $B_i$, that is, the step whereby the edge labels of $G'_i$ are transformed into the node labels of $B_i$. A node $X_a$ in $G'_i$ or $B_i$ has $|\Pi_{X_a}|$ parents, and for each of the $2^{|\Pi_{X_a}|}$ possible joint value assignments to those parents, say the value assignment $\boldsymbol{\pi}_{X_a}$, the conditional probability $p(0 \mid \boldsymbol{\pi}_{X_a})$ (or, equivalently, $p(1 \mid \boldsymbol{\pi}_{X_a})$) must be provided as part of the label of $X_a$. Evidently, requiring such an exponentially large number of label components is impractical even for moderately complex instances of $P$ and some more efficient representation must be adopted.

Our choice on this issue has been to adopt the customary noisy-OR assumption [13], whose core in our context is the following. Let $X_a$ be a node with at least one parent, and let $X_{a_1}, \ldots, X_{a_\sigma}$ be its parents, with $\sigma = |\Pi_{X_a}|$. The assumption is that whatever causes the event $X_a = 1$ to be unaffected by the event $X_{a_k} = 1$ is independent from whatever else may cause $X_a = 1$ to be unaffected by $X_{a_l} = 1$, where $X_{a_k}$ and $X_{a_l}$ are any two distinct parents of $X_a$. If we let $\boldsymbol{\pi}^k_{X_a}$ denote the joint value assignment to the parents of $X_a$ that sets $X_{a_k} = 1$

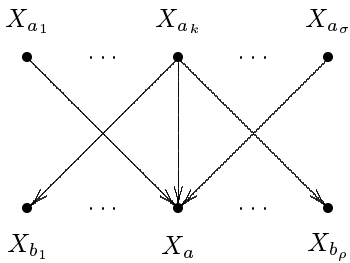

Figure 1: The surroundings of variable $X_a$ that are relevant to (5).

and all other parents to 0, then clearly the noisy-OR assumption amounts to

$$p(0 \mid \boldsymbol{\pi}_{X_a}) = \prod_{\substack{k=1 \\ X_{a_k}=1}}^{\sigma} p(0 \mid \boldsymbol{\pi}_{X_a}^k), \qquad (4)$$

where the $X_{a_k} = 1$ condition indicates that the product ranges over the parents of $X_a$ that have value 1 in $\boldsymbol{\pi}_{X_a}$. By (4), only the $\sigma$ conditional probabilities $p(0 \mid \boldsymbol{\pi}_{X_a}^1), \ldots, p(0 \mid \boldsymbol{\pi}_{X_a}^\sigma)$ need to be specified: the remaining $2^{|\Pi_{X_a}|} - \sigma$ ones can be easily computed as they become necessary.

For $1 \leq k \leq \sigma$, the following is how we obtain the value of $p(0 \mid \boldsymbol{\pi}_{X_a}^k)$ for use in $B_i$. Let $X_{b_1}, \ldots, X_{b_\rho}$ be the children of $X_{a_k}$ (including, of course, $X_a$). We then let

$$p(0 \mid \boldsymbol{\pi}_{X_a}^k) = 1 - \frac{f_{a_k a}}{f_{a_k b_1} + \cdots + f_{a_k b_\rho}}. \qquad (5)$$

In words, what (5) is saying is that $p(1 \mid \boldsymbol{\pi}_{X_a}^k)$ is, of all the times the execution of $P$ goes through basic block $a_k$, the fraction in which it proceeds directly to basic block $a$. An illustration depicting the variables involved in this process is given in Figure 1.

All we are left to do is then to handle the case in which $X_a$ has no parent and therefore needs a prior probability. This case includes the single source inherited by $B_i$ from $G_i$ and all the ones we inserted artificially when severing back edges during the transformation of $G_i$ into $G'_i$. However, as will become apparent in Section 3, all such prior probabilities are irrelevant and we need not worry about them. This is so because the source that represents the initial basic block is an evidence (so its value never changes) and all the other sources are treated, during the solution of the Bayesian network, in a somewhat unorthodox way intended to ensure that their values are consistent with those of the sinks that were created along with them.

## 2.3 The history model

For $i > 1$ (that is, after $P$ has been run at least once), the history model of the first $i-1$ executions is a Bayesian network, denoted by $H_{i-1}$, which incorporates probabilistic knowledge on the occurrence of the basic blocks of $P$ as it is executed. We now describe how to incorporate the probabilistic knowledge that $B_i$ embodies about the $i$th execution of $P$ into $H_{i-1}$ so that $H_i$, a new Bayesian network now incorporating information on the first $i$ executions, can be obtained.

In order to achieve the combination of $H_{i-1}$ and $B_i$ into $H_i$, first we must ensure that the two Bayesian networks we start with have the same node and edge sets. This can be achieved simply by first determining the union of the two node sets and the union of the two edge sets, and then enlarging each node set to make it equal the union of the node sets, then similarly for the edge sets. The only problem with this is that it leaves some node labels incomplete, i.e., in both $H_{i-1}$ and $B_i$ there may be a non-source variable $X$ with less than $|\Pi_X|$ conditional probabilities specified for it. Each missing probability is a probability that $X = 0$ given that a newly added parent has value 1 and all others value 0. What we do in these cases is to set all missing probabilities to a small value $\epsilon \in (0, 1)$.[1]

We may then henceforth assume that $H_{i-1}$ and $B_i$ have the same node and edge set, and also that they have labels completely specified within the noisy-OR assumption for all non-source nodes. These shared node and edge sets are also the node and edge sets of the resulting Bayesian network, $H_i$. Let $\mathbf{X}$ denote this common node set and $\mathbf{x}$ stand for a joint value assignment to all the variables in $\mathbf{X}$.

We would like, ideally, to obtain the node labels of $H_i$ in such a way as to ensure that the resulting joint distribution over $\mathbf{X}$ were the (normalized) geometric average of the two source joint distributions, i.e., those that correspond to $H_{i-1}$ and $B_i$. The geometric average of two distributions seems only natural in the Bayesian-network context, since it involves products of probabilities and such products already lie at the core of any analysis of Bayesian networks (cf. (1)). So if $p_i$ is the probability distribution for $B_i$ and $q_i$ the distribution for $H_i$, then we would aim at having, with $\alpha_i \in (0, 1)$ and $n_i = |\mathbf{X}|$,

$$q_i(\mathbf{x}) = \frac{q_{i-1}(\mathbf{x})^{1-\alpha_i} p_i(\mathbf{x})^{\alpha_i}}{\sum_{\mathbf{x}' \in \{0,1\}^{n_i}} q_{i-1}(\mathbf{x}')^{1-\alpha_i} p_i(\mathbf{x}')^{\alpha_i}} \tag{6}$$

for all $\mathbf{x} \in \{0, 1\}^{n_i}$. And in fact it is easy to demonstrate that (6) is achieved if it is also achieved at the node-label level, that is, if

$$q_i(x \mid \boldsymbol{\pi}_X) = \frac{q_{i-1}(x \mid \boldsymbol{\pi}_X)^{1-\alpha_i} p_i(x \mid \boldsymbol{\pi}_X)^{\alpha_i}}{\sum_{x' \in \{0,1\}} q_{i-1}(x' \mid \boldsymbol{\pi}_X)^{1-\alpha_i} p_i(x' \mid \boldsymbol{\pi}_X)^{\alpha_i}} \tag{7}$$

for all $X \in \mathbf{X}$, all $x \in \{0, 1\}$, and all $\boldsymbol{\pi}_X \in \{0, 1\}^{|\Pi_X|}$.

Let us digress briefly to outline the main argument of this demonstration. If we assume that (7) holds as stated, then we obtain, for all $\mathbf{x} \in \{0, 1\}^{n_i}$ and starting with an application of (1),

$$q_i(\mathbf{x}) = \prod_{X \in \mathbf{X}} q_i(x \mid \boldsymbol{\pi}_X) \tag{8}$$

$$= \frac{\prod_{X \in \mathbf{X}} q_{i-1}(x \mid \boldsymbol{\pi}_X)^{1-\alpha_i} p_i(x \mid \boldsymbol{\pi}_X)^{\alpha_i}}{\prod_{X \in \mathbf{X}} \sum_{x' \in \{0,1\}} q_{i-1}(x' \mid \boldsymbol{\pi}_X)^{1-\alpha_i} p_i(x' \mid \boldsymbol{\pi}_X)^{\alpha_i}}. \tag{9}$$

---

[1] We note that it is critical that $\epsilon$ be a strictly positive value. Setting such probabilities to 0 disrupts the fundamental nature of a Bayesian network as a Markov (or, equivalently, a Gibbs) random field, in which case all the theory that underlies the optimization process summarized by (3) crumbles [3].

Rewriting the denominator yields

$$\frac{\prod_{X \in \mathbf{X}} q_{i-1}(x \mid \boldsymbol{\pi}_X)^{1-\alpha_i} p_i(x \mid \boldsymbol{\pi}_X)^{\alpha_i}}{\sum_{\mathbf{x}' \in \{0,1\}^{n_i}} \prod_{X \in \mathbf{X}} q_{i-1}(x' \mid \boldsymbol{\pi}_X)^{1-\alpha_i} p_i(x' \mid \boldsymbol{\pi}_X)^{\alpha_i}}, \tag{10}$$

where $x'$ is the value of $X$ in $\mathbf{x}'$. By (1), this leads to (6).

The problem is that the $q_i(x \mid \boldsymbol{\pi}_X)$ of (7) is in general not compliant with the noisy-OR assumption we made in Section 2.2 even if $q_{i-1}(x \mid \boldsymbol{\pi}_X)$ and $p_i(x \mid \boldsymbol{\pi}_X)$ are. In order to see this, we assume the latter and rewrite (7) using the notation of Section 2.2 with $\sigma = |\Pi_X|$; by (4), we get, for instance for $x = 0$,

$$q_i(0 \mid \boldsymbol{\pi}_X) = \frac{\prod_{\substack{k=1 \\ X_k=1}}^{\sigma} q_{i-1}(0 \mid \boldsymbol{\pi}_X^k)^{1-\alpha_i} p_i(0 \mid \boldsymbol{\pi}_X^k)^{\alpha_i}}{\sum_{x' \in \{0,1\}} q_{i-1}(x' \mid \boldsymbol{\pi}_X)^{1-\alpha_i} p_i(x' \mid \boldsymbol{\pi}_X)^{\alpha_i}}, \tag{11}$$

where the denominator can also be rewritten:

$$\prod_{\substack{k=1 \\ X_k=1}}^{\sigma} q_{i-1}(0 \mid \boldsymbol{\pi}_X^k)^{1-\alpha_i} p_i(0 \mid \boldsymbol{\pi}_X^k)^{\alpha_i}$$
$$+ \left(1 - \prod_{\substack{k=1 \\ X_k=1}}^{\sigma} q_{i-1}(0 \mid \boldsymbol{\pi}_X^k)\right)^{1-\alpha_i} \left(1 - \prod_{\substack{k=1 \\ X_k=1}}^{\sigma} p_i(0 \mid \boldsymbol{\pi}_X^k)\right)^{\alpha_i}. \tag{12}$$

Clearly, for noisy-OR compliance we should have

$$q_i(0 \mid \boldsymbol{\pi}_X) = \prod_{\substack{k=1 \\ X_k=1}}^{\sigma} \frac{q_{i-1}(0 \mid \boldsymbol{\pi}_X^k)^{1-\alpha_i} p_i(0 \mid \boldsymbol{\pi}_X^k)^{\alpha_i}}{\sum_{x' \in \{0,1\}} q_{i-1}(x' \mid \boldsymbol{\pi}_X^k)^{1-\alpha_i} p_i(x' \mid \boldsymbol{\pi}_X^k)^{\alpha_i}}$$
$$= \frac{\prod_{\substack{k=1 \\ X_k=1}}^{\sigma} q_{i-1}(0 \mid \boldsymbol{\pi}_X^k)^{1-\alpha_i} p_i(0 \mid \boldsymbol{\pi}_X^k)^{\alpha_i}}{\prod_{\substack{k=1 \\ X_k=1}}^{\sigma} \sum_{x' \in \{0,1\}} q_{i-1}(x' \mid \boldsymbol{\pi}_X^k)^{1-\alpha_i} p_i(x' \mid \boldsymbol{\pi}_X^k)^{\alpha_i}}, \tag{13}$$

but the two denominators are not in general equal.

The inescapable conclusion is then that we must choose between the concise node-label representations afforded by the noisy-OR assumption and achieving (6) through (7). Given our application domain, in which variables with hundreds of parents do occur, the ability to represent node labels parsimoniously is absolutely essential. We then choose the first option while the second one remains an ideal to be approximated. Having opted for conciseness, it then suffices to apply the geometric-average rule of (7) to the $|\Pi_X|$ conditional probabilities of each and every variable $X \in \mathbf{X}$ that is not a source (prior probabilities, recall, are in our context needless).

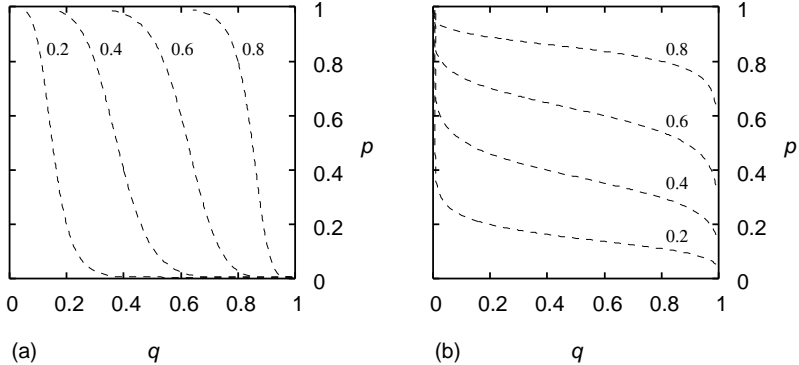

Figure 2: Contour plots of $r = q^{1-\alpha}p^\alpha / \left(q^{1-\alpha}p^\alpha + (1-q)^{1-\alpha}(1-p)^\alpha\right)$ with $0 \le p, q \le 1$ for $\alpha = 0.2$ (a) and $\alpha = 0.8$ (b). As $\alpha$ is increased from 0.2 in (a) to 0.8 in (b), $r$ becomes more sensitive to the value of $p$ than to the value of $q$.

It now remains for us to find a suitable value for $\alpha_i$. The strategy we use is based on the following general premise. Suppose we can devise an ideal value, call it $\alpha_0$, for the mixture of the two Bayesian networks. This can be done, for instance, by running $P$ a number of times on a randomly chosen sequence of inputs, each time with a different candidate value for $\alpha_0$, and at the end selecting the value that yields the smallest overall running time. The chosen $\alpha_0$ can then be used as a sort of threshold: after $P$ is run on $I_i$ and $B_i$ is obtained, we check its running time against some average of the running times of $P$ on the previous $i-1$ inputs; if smaller we select a value for $\alpha_i$ that is smaller than $\alpha_0$, and correspondingly if it is larger. What this is doing, since we are dealing with geometric averages of numbers below 1, is to let executions with comparatively larger running times weigh more in the history model than executions with comparatively smaller running times (essentially, the probability that gets raised to the smallest exponent yields, if large enough, the result that is nearest 1 and therefore affects the geometric average the least—cf. Figure 2 for a clarifying illustration).

Now for the details. Let $t_1, t_2, \ldots$ be the running times of $P$ on $I_1, I_2, \ldots$, respectively. Let $T_{i-1}$ be the average, weighted by normalized versions of $\alpha_1, \ldots, \alpha_{i-1}$, of the first $i-1$ running times, that is,

$$T_{i-1} = \frac{\alpha_1 t_1 + \cdots + \alpha_{i-1} t_{i-1}}{\alpha_1 + \cdots + \alpha_{i-1}}. \tag{14}$$

The value we use for $\alpha_i$ is then

$$\alpha_i = \frac{1}{1 + \left(\frac{1-\alpha_0}{\alpha_0}\right) e^{-\gamma(t_i - T_{i-1})}}, \tag{15}$$

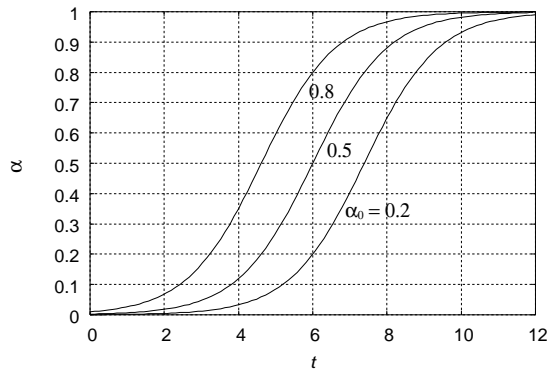

Figure 3: Plots of $\alpha = \left(1 + e^{T-t}(1-\alpha_0)/\alpha_0\right)^{-1}$ with $T = 6$ for $\alpha_0 = 0.2, 0.5, 0.8$.

where $\gamma > 0$ is a parameter. The functional dependency in (15) has a sigmoidal form and maps $t_i$ into the interval $[0, 1]$. It yields $\alpha_i = \alpha_0$ when $t_i = T_{i-1}$, that is, when $P$ runs on the $i$th input as fast as it has on average; smaller values of $t_i$ bring $\alpha_i$ closer to 0, larger values bring it closer to 1. The function's steepness around $t_i = T_{i-1}$ is controlled by the $\gamma$ parameter. Illustrations with $\gamma = 1$ are given in Figure 3.

All of our discussion concerning the evolution of the history model holds, of course, for $i > 1$. For $i = 1$, no previous history model exists and $B_1$ simply becomes $H_1$ while we let $T_1 = t_1$. In addition, determining $\alpha_1$ for later use requires that $T_0$ be known as well; there are various possibilities, one being to take $T_0$ as the average running time of $P$ during the initial experiments that yield the value of $\alpha_0$.

## 3 The overall strategy

The following is a summary of our strategy in this paper. It provides the main steps to be followed as the program $P$ is run, in succession, on the inputs $I_1, I_2, \ldots$. We assume that running $P$ on $I_i$ for $i \geq 1$ automatically produces the Bayesian network $B_i$ as explained in Section 2.2 and also yields the measure $t_i$ for the running time of $P$ on $I_i$. This running time is assumed to exclude all the instrumentation effort that creates $B_i$. We also assume that the value of $\alpha_0$ is known from previous experimentation with $P$ on a random assortment of inputs, and furthermore that the average running time of $P$ during the experiments is recorded as $T_0$.

1. Run $P$ on $I_1$, then do:

    (a) Determine $\alpha_1$ from (15).
    (b) Let $T_1 = t_1$.

(c) Duplicate $B_1$ to yield $H_1$.

(d) Solve $H_1$ for the most likely joint occurrence of basic blocks, then reorder the basic blocks of $P$ accordingly.

2. For $i > 1$, run $P$ on $I_i$, then do:

   (a) Determine $\alpha_i$ from (15).

   (b) Let
   $$T_i = \frac{\alpha_1 t_1 + \cdots + \alpha_i t_i}{\alpha_1 + \cdots + \alpha_i}.$$

   (c) Obtain the node and edge set of $H_i$ as the union, respectively, of the node and edge sets of $H_{i-1}$ and $B_i$. Let $\mathbf{X}$ be the node set of $H_i$.

   (d) For $X \in \mathbf{X}$, do the following if $X$ is not a source. Let $\sigma = |\Pi_X|$ and $\Pi_X = \{X_1, \ldots, X_\sigma\}$. Let also $\boldsymbol{\pi}_X^k$, with $1 \leq k \leq \sigma$, be the joint value assignment to the variables in $\Pi_X$ that assigns 0 to all variables except $X_k$, which receives value 1. For $k = 1, \ldots, \sigma$, let
   $$q_i(0 \mid \boldsymbol{\pi}_X^k) = \frac{q_{i-1}(0 \mid \boldsymbol{\pi}_X^k)^{1-\alpha_i} p_i(0 \mid \boldsymbol{\pi}_X^k)^{\alpha_i}}{Z_i^k},$$
   where
   $$\begin{aligned}Z_i^k &= q_{i-1}(0 \mid \boldsymbol{\pi}_X^k)^{1-\alpha_i} p_i(0 \mid \boldsymbol{\pi}_X^k)^{\alpha_i} \\ &\quad + \left(1 - q_{i-1}(0 \mid \boldsymbol{\pi}_X^k)\right)^{1-\alpha_i} \left(1 - p_i(0 \mid \boldsymbol{\pi}_X^k)\right)^{\alpha_i}.\end{aligned}$$
   If either $q_{i-1}(0 \mid \boldsymbol{\pi}_X^k)$ or $p_i(0 \mid \boldsymbol{\pi}_X^k)$ is missing (because $X_k$ is not a parent of $X$ in both $H_{i-1}$ and $B_i$), then assume a small value $\epsilon \in (0, 1)$ for it.

   (e) Solve $H_i$ for the most likely joint occurrence of basic blocks, then reorder the basic blocks of $P$ accordingly.

Solving the history models in Steps 1(d) and 2(e) can be achieved, for example, by the variation of stochastic simulation mentioned in Section 2.1. During the simulation, the variable that corresponds to the initial basic block is treated as an evidence, that is, its value remains fixed at 1 at all times.

All other variables have their values updated regularly according to the probability prescribed in (3), but the following special precaution is taken when updating the source-sink pairs of variables created as back edges are severed during the construction of the execution model. If $X_{a'}$ and $X_{b'}$ are, respectively, such a source and sink, then $X_{a'}$ is never updated directly but rather has its value copied from that of $X_{b'}$ whenever $X_{b'}$ is updated. This is intended to ensure the semantic consistency that the creation of the two variables implicitly suggests as desirable.

The reordering of $P$'s basic blocks in the same two Steps 1(d) and 2(e) involves examining all the variables that have value 1 in the global joint value assignment obtained as solution of the history model. This assignment is an approximation of an $\mathbf{x}^*$ for which $q_i(\mathbf{x}^*)$ is maximum given the evidence corresponding to the execution of the initial basic block. In $\mathbf{x}^*$, we expect variables with value 1 to constitute a unique directed path in the history model whose first variable corresponds to the initial basic block, provided we allow jumps between variables like the $X_{b'}$ and $X_{a'}$ above to be included in the path. As we discuss in Section 4, this expectation has in practice been verified for the approximations of $\mathbf{x}^*$ that we obtain as well. Reordering the basic blocks is then simply a matter of placing the basic blocks of this directed path in a position inside $P$ that ensures they are the first to be loaded into memory for execution on the next input.

Note, finally, that node labels for both the execution models, via (5), and the history models, via Step 2(d), are stored concisely according to the noisy-OR assumption. By (4), all the conditional probabilities not explicitly stored may be obtained readily when needed during the simulation.

## 4 Experimental results

We have conducted extensive experiments to assess the performance of the strategy summarized in Steps 1 and 2 of Section 3, henceforth referred to as the Bayesian-network approach. Our goal has been twofold: first, to verify the approach's ability to provide better running times as a program is repeatedly run on the same input, possibly with the intervention of other inputs; secondly, to compare the running times under the Bayesian-network approach with those obtained under the gcc level-3 optimization (with no further code reordering) or Pettis-Hansen reordering [11] (as implemented in PLTO [14]). In the remainder of the paper, we refer to the latter two strategies concisely by the epithets O3 and PH, respectively.

The PH strategy can be viewed as operating precisely on the $G_i$ graph of Section 2. In essence, what it does is to repeatedly concatenate basic-block chains greedily based on the counts that label the graph's edges. To this end, it first lets every basic block be a chain, and then proceeds by examining the edges that connect the end of a chain to the beginning of another and selecting the one that has the greatest label to join its end chains. When chains can no longer be joined, they are placed in a relative order that favors the most frequently taken branches. The program's basic blocks are then reordered accordingly. PLTO, including as it does the functionality to do basic-block reordering from edge-labeled graphs like $G_i$, provides a convenient framework for implementing not only the PH strategy (which it does by default) but also the basic-block reordering prescribed by our Bayesian-network approach (which we lead it to achieve, as discussed below).

In addition to this use of PLTO to achieve the reordering of basic blocks, we also use it as part of the procedure to generate the graph $G_i$, as it already implements a considerable portion of the profiling functionality that is necessary to build that graph. However, PLTO does this profiling separately for each procedure and does not provide the frequency counts that correspond to returns from executing procedures. But in $G_i$ every node and edge must be properly placed (and, in the case of edges, labeled), so in essence we do the following addition to the processing of PLTO. Let $a$ be a basic block through which a certain procedure is called, and let $b$ be the basic block that follows after the procedure is executed. PLTO provides the edge labels inside the procedure's code but links $a$ directly to $b$ along with the label $f_{ab}$. But what we need in $G_i$, instead, is an edge directed from $a$ to the procedure's entry basic block, and also an edge directed from the procedure's exit basic block to $b$. We then create each of these edges with label $f_{ab}$ and eliminate $(a \rightarrow b)$.

Table 1: Input distribution.

| Program | Reference | Train | Test | Reduced |
|---|---|---|---|---|
| bzip2 | 0–2 | 3 | 4 | 5–7 |
| crafty | 0 | 1 | 2 | 3–5 |
| gap | 0 | 1 | 2 | 3–5 |
| gcc | 0–4 | 5 | 6 | 7, 8 |
| gzip | 0–4 | 5 | 6 | 7–21 |
| mcf | 0 | 1 | 2 | 3, 4 |
| parser | 0 | 1 | 2 | 3–5 |
| twolf | 0 | 1 | 2 | 3, 4 |
| vortex | 0–2 | 3 | 4 | 5–7 |
| vpr | 0, 1 | 2, 3 | 4, 5 | 6–9 |

We concentrated our experiments on the SPEC programs listed in Table 1.[2] For each program, the inputs that appear in the suite's Reference, Train, Test, and Reduced sets are numbered sequentially from 0 as indicated in the table. Within each of the four sets, our numbering follows the same order as used for the suite's files.

We used for all our experiments an AMD Athlon running at 1.66 GHz with a 256-Kbyte level-2 cache and 256 Mbytes of main memory. We used the RedHat 7.3 Linux operating system (kernel version 2.4.18-3) and version 2.96 of gcc (always with the level-3 optimization option). Every running time we report is expressed in seconds and refers, for each program on each input, to the middle time of three runs (i.e., the one that remains after discarding the lowest and highest times).

Let $n_P$ denote the number of distinct inputs to a program $P$ from the SPEC suite (in Table 1, inputs are then numbered from 0 through $n_P - 1$). Our methodology for experimentation on $P$ has been to apply the Bayesian-network approach to the sequence $I_1, \ldots, I_{6n_P}$ of inputs to $P$ generated randomly in such a way that each of the $n_P$ inputs appears exactly six times in the sequence. In order to apply the Bayesian-network approach to this sequence via Steps 1 and 2 of Section 3, we first chose the values of $\alpha_0$ and $T_0$ as indicated in that section,[3] and then proceeded to one of the two steps on each of the $6n_P$ inputs with $\gamma = 1$.

---

[2] We have omitted eon and perlbmk from our experiments because there seems to be some incompatibility with the PLTO version that is current as we write. But the conclusions we draw in the sequel appear to be well supported by the programs we did consider, so we believe these two omissions to be essentially harmless.

Whenever applying (3) for solving a history model by stochastic simulation, we let $T$ vary from $T = 10^4$ down to the first value below 1 that could be obtained by letting each new value for $T$ be 98% of its predecessor. It then follows that $T$ went through $\lceil 1 - \ln 10^4 / \ln 0.98 \rceil = 457$ continually decreasing values. For each value, each non-source variable of the model was updated exactly once according to the probability in (3). At the end, searching through the variables for those with value 1 consistently yielded, in all occasions, the desired directed path from the initial basic block described in Section 3.

In our experiments, this is where our use of PLTO's functionality once again comes in. Specifically, we revert to the $G_i$ graph and modify its edge labels as follows before feeding it to PLTO. First, edges that do not appear in the directed path that solves the history model have their labels set to 0. Then we scan the edges of that directed path, starting at the initial basic block, and change all edge labels by assigning a large label (at least $n$) to the firsts edge encountered, this first label minus 1 to the second edge, and so on. Indirectly, this necessarily leads PLTO's default strategy (the PH strategy) to reorder the basic blocks as desired.

Our results on the programs of Table 1 are summarized in the plots of Figure 4. For each of the ten programs, the figure contains two sets of plots displayed side by side. The plots on the left refer to inputs on which the program is considerably slower than on those whose plots appear on the right (with only two exceptions, `gcc` and `gzip`, plots on the left correspond exactly to the inputs in the Reference set). The two plot sets for each program share the same abscissae representing the sequence numbers of the various input instances of that program in the randomly generated sequence of inputs. Each plot comprises six points connected by a line, each point referring to a different occurrence of the same input in the sequence.

Dividing the plots for each program in this manner does more than simply solve the scale problem on the ordinate axis: at least qualitatively (and also quantitatively, provided one bears in mind the differences in scale between the left- and right-hand sides), the division highlights the fact that the program's performance on inputs on the left (those for which running times are larger) tends to improve noticeably as the sequence of inputs is played, particularly when the same input occurs in the sequence with little or no occurrence of other intervening inputs. This is to be contrasted with what happens to the inputs on the right (the ones for which running times are smaller). With only a few exceptions, on these inputs the program tends to perform in a relatively unaltered way, yielding practically the same running time at all encounters with the same input. Interestingly, the exceptions are precisely those inputs for which the running times stand out within the sets on the right-hand side, that is, those that require substantially larger running times than the other inputs in their sets. This pattern of behavior is, of course, what we aimed at with the design summarized in Section 3.

---

[3]We found $\alpha_0 = 0.8$ to be satisfactory regardless of $P$, and also the particular value assigned to $T_0$ to be practically immaterial.

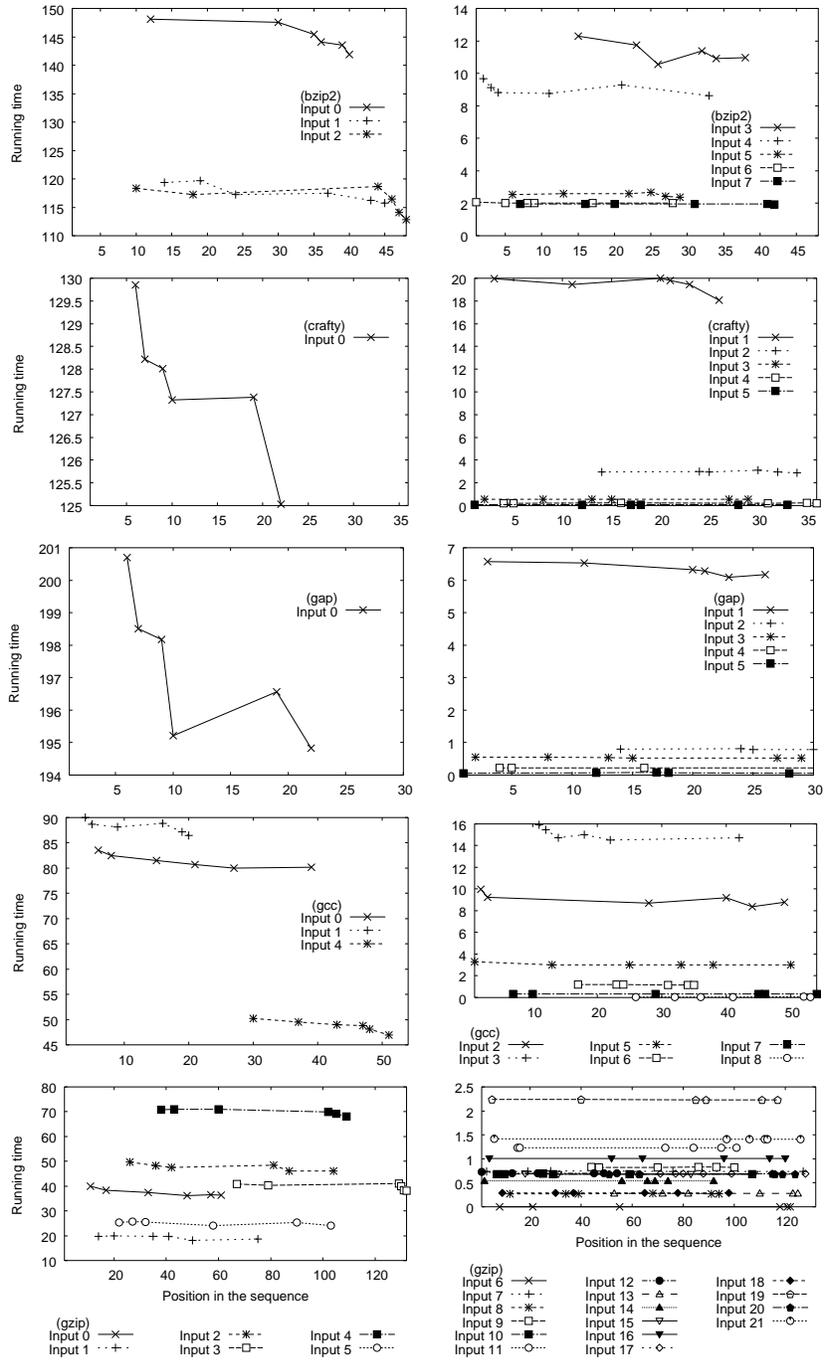

Figure 4: Performance evolution for the SPEC programs of Table 1.

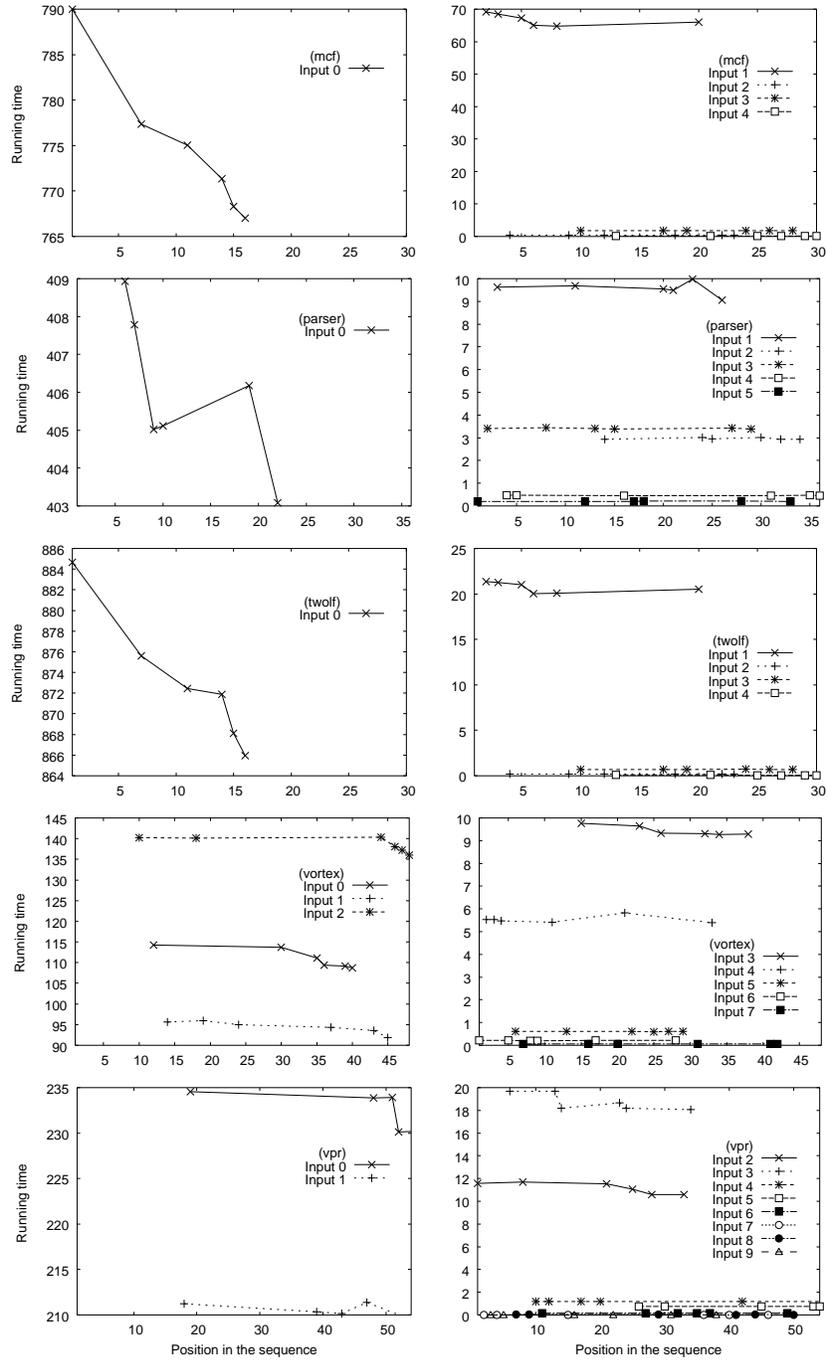

Figure 4: (Continued).

Table 2: Performance data for `bzip2`.

| Input | $t_\text{last}$ | $t_\text{first}$ | $g_\text{first}$ | $t_\text{O3}$ | $g_\text{O3}$ | $t_\text{PH}$ | $g_\text{PH}$ |
|---|---|---|---|---|---|---|---|
| 0 | 141.87 | 148.13 | 4.23 | 148.91 | 4.73 | 148.11 | 4.21 |
| 1 | 115.76 | 119.34 | 3.00 | 122.51 | 5.51 | 119.47 | 3.11 |
| 2 | 112.87 | 118.34 | 4.62 | 120.76 | 6.53 | 118.03 | 4.37 |
| 3 | 10.97 | 12.28 | 10.67 | 12.07 | 9.11 | 11.98 | 8.43 |
| 4 | 8.63 | 9.67 | 10.75 | 9.77 | 11.67 | 9.63 | 10.38 |
| 5 | 2.38 | 2.54 | 6.30 | 2.65 | 10.18 | 2.59 | 8.11 |
| 6 | 2.02 | 2.07 | 2.42 | 2.07 | 2.42 | 2.02 | 0.00 |
| 7 | 1.94 | 1.95 | 0.51 | 2.04 | 4.90 | 1.97 | 1.52 |

Table 3: Performance data for `crafty`.

| Input | $t_\text{last}$ | $t_\text{first}$ | $g_\text{first}$ | $t_\text{O3}$ | $g_\text{O3}$ | $t_\text{PH}$ | $g_\text{PH}$ |
|---|---|---|---|---|---|---|---|
| 0 | 125.03 | 129.85 | 3.71 | 133.25 | 6.16 | 126.21 | 0.94 |
| 1 | 18.06 | 19.98 | 9.61 | 20.12 | 10.24 | 19.30 | 6.42 |
| 2 | 2.88 | 2.97 | 3.03 | 3.08 | 6.49 | 2.91 | 1.03 |
| 3 | 0.56 | 0.56 | 0.00 | 0.58 | 3.45 | 0.55 | −1.82 |
| 4 | 0.21 | 0.21 | 0.00 | 0.23 | 8.70 | 0.21 | 0.00 |
| 5 | 0.07 | 0.07 | 0.00 | 0.07 | 0.00 | 0.07 | 0.00 |

In Tables 2 through 11, one for each of the ten programs, we provide data that help interpret the plots of Figure 4, and also data for comparing our Bayesian-network approach with the O3 and PH strategies. In each table, the $t_\text{last}$ column gives, for each input, the program's running time on the last (the sixth) occurrence of that input in the randomly generated sequence of inputs, while $t_\text{first}$ gives the running time on the input's first occurrence. Similarly, $t_\text{O3}$ and $t_\text{PH}$ refer to the running times, respectively, of the O3 and PH versions of the program on that same input. Note that $t_\text{first}$ and $t_\text{O3}$ are expected to be practically the same for the input that happens to be the first in the randomly generated sequence (for this input, and recalling that all compilations do level-3 optimization, the O3 strategy is indistinguishable from ours).

The remaining three columns in the tables give the final percent gain of the Bayesian-network approach over the first encounter with each input, and also over the O3 and PH versions. These three gains are defined, respectively, as $g_\text{first} = 100(t_\text{first} - t_\text{last})/t_\text{first}$, $g_\text{O3} = 100(t_\text{O3} - t_\text{last})/t_\text{O3}$, and $g_\text{PH} = 100(t_\text{PH} - t_\text{last})/t_\text{PH}$. With very rare exceptions, positive gains dominate the ten tables and indicate a superior performance, by up to 12.61% in the case that relates to the first execution on the input in question, 14.50% in the case of O3, and 12.96% in the PH case, of the Bayesian-network approach. The figures for $g_\text{first}$ clearly corroborate our conclusions when analyzing Figure 4. Also, gains over O3 tend to surpass those over PH, once again with rare exceptions.

Table 4: Performance data for gap.

| Input | $t_{\text{last}}$ | $t_{\text{first}}$ | $g_{\text{first}}$ | $t_{\text{O3}}$ | $g_{\text{O3}}$ | $t_{\text{PH}}$ | $g_{\text{PH}}$ |
|---|---|---|---|---|---|---|---|
| 0 | 194.83 | 200.70 | 2.92 | 203.78 | 4.39 | 200.98 | 3.06 |
| 1 | 6.17 | 6.57 | 6.09 | 6.49 | 4.93 | 6.41 | 3.75 |
| 2 | 0.78 | 0.80 | 2.50 | 0.83 | 6.02 | 0.79 | 1.27 |
| 3 | 0.52 | 0.55 | 5.45 | 0.56 | 7.14 | 0.53 | 1.89 |
| 4 | 0.22 | 0.22 | 0.00 | 0.22 | 0.00 | 0.24 | 8.33 |
| 5 | 0.06 | 0.06 | 0.00 | 0.06 | 0.00 | 0.06 | 0.00 |

Table 5: Performance data for gcc.

| Input | $t_{\text{last}}$ | $t_{\text{first}}$ | $g_{\text{first}}$ | $t_{\text{O3}}$ | $g_{\text{O3}}$ | $t_{\text{PH}}$ | $g_{\text{PH}}$ |
|---|---|---|---|---|---|---|---|
| 0 | 80.15 | 83.53 | 4.05 | 85.89 | 6.68 | 83.31 | 3.79 |
| 1 | 86.43 | 89.98 | 3.95 | 92.82 | 6.89 | 89.78 | 3.73 |
| 2 | 8.78 | 9.98 | 12.02 | 10.27 | 14.50 | 8.95 | 1.90 |
| 3 | 14.73 | 15.90 | 7.55 | 15.92 | 7.66 | 14.97 | 1.80 |
| 4 | 46.98 | 50.20 | 6.41 | 52.02 | 9.69 | 50.82 | 7.56 |
| 5 | 3.01 | 3.29 | 8.51 | 3.29 | 8.51 | 3.01 | 0.00 |
| 6 | 1.16 | 1.21 | 4.13 | 1.31 | 11.45 | 1.16 | 0.00 |
| 7 | 0.32 | 0.32 | 0.00 | 0.35 | 8.57 | 0.32 | 0.00 |
| 8 | 0.06 | 0.06 | 0.00 | 0.06 | 0.00 | 0.06 | 0.00 |

## 5 Concluding remarks

We have in this paper introduced a new approach to improving the usage of the instruction memory. Our approach is probabilistic in nature and has two main ingredients. The first ingredient is what we call the execution model and concerns each individual execution of a given program on a given input. The execution model is a Bayesian network whose node labels are built from a trace recorded as the program is run on the input. The second ingredient is what we call the history model. It too is a Bayesian network, one that is now focused on a given program as it runs on a varied stream of inputs: after the program is run on a new input, the resulting execution model is incorporated into the history model by a technique that updates node labels to the geometric average of two corresponding labels, one from the current history model, the other from the new execution model.

The effect of this continual updating of the history model by data resulting from running the program on new inputs is that, for each new input, the actual code to be executed can take into account all the knowledge stored in the history model. The way this is achieved is by reordering basic blocks for use on the next input whenever the history model gets updated. This reordering is done in such a way that the program's basic blocks that according to the history model are the most likely to be executed are the ones to occupy the highest levels of the instruction-memory hierarchy.

Table 6: Performance data for gzip.

| Input | $t_{\text{last}}$ | $t_{\text{first}}$ | $g_{\text{first}}$ | $t_{\text{O3}}$ | $g_{\text{O3}}$ | $t_{\text{PH}}$ | $g_{\text{PH}}$ |
|---|---|---|---|---|---|---|---|
| 0 | 36.38 | 39.87 | 8.75 | 40.77 | 10.77 | 40.13 | 9.34 |
| 1 | 18.65 | 19.83 | 5.95 | 20.72 | 9.99 | 19.89 | 6.23 |
| 2 | 46.11 | 49.76 | 7.34 | 51.14 | 9.84 | 50.06 | 7.89 |
| 3 | 38.11 | 40.87 | 6.75 | 42.52 | 10.37 | 41.34 | 7.81 |
| 4 | 68.08 | 70.76 | 3.79 | 74.26 | 8.32 | 72.23 | 5.86 |
| 5 | 24.14 | 25.30 | 4.58 | 26.96 | 10.46 | 26.56 | 9.11 |
| 6 | 0.00 | 0.00 | 0.00 | 0.00 | 0.00 | 0.00 | 0.00 |
| 7 | 0.74 | 0.74 | 0.00 | 0.76 | 2.63 | 0.74 | 0.00 |
| 8 | 0.27 | 0.27 | 0.00 | 0.29 | 6.90 | 0.27 | 0.00 |
| 9 | 0.82 | 0.83 | 1.20 | 0.85 | 3.53 | 0.84 | 2.38 |
| 10 | 0.68 | 0.68 | 0.00 | 0.70 | 2.86 | 0.68 | 0.00 |
| 11 | 1.23 | 1.23 | 0.00 | 1.27 | 3.15 | 1.25 | 1.60 |
| 12 | 0.70 | 0.73 | 4.11 | 0.73 | 4.11 | 0.71 | 1.41 |
| 13 | 0.28 | 0.28 | 0.00 | 0.30 | 6.67 | 0.28 | 0.00 |
| 14 | 0.54 | 0.54 | 0.00 | 0.57 | 5.26 | 0.54 | 0.00 |
| 15 | 0.68 | 0.69 | 1.45 | 0.72 | 5.55 | 0.69 | 1.44 |
| 16 | 1.01 | 1.01 | 0.00 | 1.05 | 3.80 | 1.02 | 0.98 |
| 17 | 0.69 | 0.69 | 0.00 | 0.71 | 2.82 | 0.70 | 1.43 |
| 18 | 0.29 | 0.29 | 0.00 | 0.30 | 3.33 | 0.29 | 0.00 |
| 19 | 2.23 | 2.24 | 0.45 | 2.32 | 3.88 | 2.27 | 1.76 |
| 20 | 0.68 | 0.68 | 0.00 | 0.71 | 4.23 | 0.69 | 1.44 |
| 21 | 1.41 | 1.42 | 0.71 | 1.49 | 5.37 | 1.44 | 2.08 |

Table 7: Performance data for mcf.

| Input | $t_{\text{last}}$ | $t_{\text{first}}$ | $g_{\text{first}}$ | $t_{\text{O3}}$ | $g_{\text{O3}}$ | $t_{\text{PH}}$ | $g_{\text{PH}}$ |
|---|---|---|---|---|---|---|---|
| 0 | 767.03 | 789.98 | 2.91 | 789.98 | 2.91 | 786.53 | 2.48 |
| 1 | 66.05 | 69.23 | 4.59 | 69.34 | 4.74 | 69.34 | 4.74 |
| 2 | 0.36 | 0.37 | 2.70 | 0.37 | 2.70 | 0.36 | 0.00 |
| 3 | 1.85 | 1.85 | 0.00 | 1.86 | 0.54 | 1.86 | 0.54 |
| 4 | 0.13 | 0.13 | 0.00 | 0.13 | 0.00 | 0.13 | 0.00 |

Table 8: Performance data for parser.

| Input | $t_{\text{last}}$ | $t_{\text{first}}$ | $g_{\text{first}}$ | $t_{\text{O3}}$ | $g_{\text{O3}}$ | $t_{\text{PH}}$ | $g_{\text{PH}}$ |
|---|---|---|---|---|---|---|---|
| 0 | 403.08 | 408.93 | 1.45 | 414.80 | 2.84 | 408.07 | 1.24 |
| 1 | 9.06 | 9.62 | 5.82 | 9.80 | 7.55 | 9.63 | 5.92 |
| 2 | 2.94 | 2.93 | −0.34 | 3.09 | 4.85 | 3.01 | 2.33 |
| 3 | 3.38 | 3.41 | 0.88 | 3.45 | 2.03 | 3.41 | 0.88 |
| 4 | 0.46 | 0.48 | 4.17 | 0.48 | 4.17 | 0.46 | 0.00 |
| 5 | 0.20 | 0.20 | 0.00 | 0.20 | 0.00 | 0.20 | 0.00 |

Table 9: Performance data for twolf.

| Input | $t_{\text{last}}$ | $t_{\text{first}}$ | $g_{\text{first}}$ | $t_{\text{O3}}$ | $g_{\text{O3}}$ | $t_{\text{PH}}$ | $g_{\text{PH}}$ |
|---|---|---|---|---|---|---|---|
| 0 | 865.97 | 884.64 | 2.11 | 884.64 | 2.11 | 875.21 | 1.06 |
| 1 | 20.53 | 21.36 | 3.89 | 21.76 | 5.65 | 21.54 | 4.69 |
| 2 | 0.20 | 0.20 | 0.00 | 0.21 | 4.76 | 0.20 | 0.00 |
| 3 | 0.71 | 0.71 | 0.00 | 0.72 | 1.39 | 0.70 | $-1.43$ |
| 4 | 0.07 | 0.08 | 12.50 | 0.08 | 12.50 | 0.08 | 12.50 |

Table 10: Performance data for vortex.

| Input | $t_{\text{last}}$ | $t_{\text{first}}$ | $g_{\text{first}}$ | $t_{\text{O3}}$ | $g_{\text{O3}}$ | $t_{\text{PH}}$ | $g_{\text{PH}}$ |
|---|---|---|---|---|---|---|---|
| 0 | 108.78 | 114.21 | 4.75 | 116.00 | 6.22 | 114.32 | 4.85 |
| 1 | 91.81 | 95.67 | 4.03 | 97.50 | 5.84 | 95.97 | 4.33 |
| 2 | 136.02 | 140.25 | 3.02 | 143.29 | 5.07 | 141.12 | 3.61 |
| 3 | 9.29 | 9.76 | 4.82 | 10.13 | 8.29 | 9.87 | 5.88 |
| 4 | 5.39 | 5.53 | 2.53 | 5.77 | 6.59 | 5.54 | 2.71 |
| 5 | 0.61 | 0.61 | 0.00 | 0.65 | 6.15 | 0.59 | $-3.39$ |
| 6 | 0.21 | 0.22 | 4.55 | 0.22 | 4.55 | 0.21 | 0.00 |
| 7 | 0.05 | 0.05 | 0.00 | 0.05 | 0.00 | 0.05 | 0.00 |

Table 11: Performance data for vpr.

| Input | $t_{\text{last}}$ | $t_{\text{first}}$ | $g_{\text{first}}$ | $t_{\text{O3}}$ | $g_{\text{O3}}$ | $t_{\text{PH}}$ | $g_{\text{PH}}$ |
|---|---|---|---|---|---|---|---|
| 0 | 226.91 | 233.47 | 2.81 | 240.89 | 2.81 | 237.21 | 4.34 |
| 1 | 206.21 | 211.91 | 2.69 | 225.39 | 8.51 | 212.52 | 2.97 |
| 2 | 10.88 | 12.45 | 12.61 | 12.45 | 12.61 | 12.50 | 12.96 |
| 3 | 18.46 | 19.58 | 5.72 | 21.05 | 12.30 | 19.76 | 6.58 |
| 4 | 1.19 | 1.18 | $-0.85$ | 1.24 | 4.03 | 1.22 | 2.46 |
| 5 | 0.73 | 0.74 | 1.35 | 0.77 | 5.19 | 0.74 | 1.35 |
| 6 | 0.17 | 0.17 | 0.00 | 0.18 | 5.56 | 0.17 | 0.00 |
| 7 | 0.00 | 0.00 | 0.00 | 0.00 | 0.00 | 0.00 | 0.00 |
| 8 | 0.02 | 0.02 | 0.00 | 0.01 | $-100.00$ | 0.01 | $-100.00$ |
| 9 | 0.00 | 0.00 | 0.00 | 0.00 | 0.00 | 0.00 | 0.00 |

Incorporating an execution model into the history model can be achieved in various ways, even if we restrict ourselves to using the geometric-average criterion. The particular way we have chosen in this paper has been to select weights for the geometric average that lets comparatively longer executions influence the history model more heavily than comparatively shorter ones. Our results on selected SPEC programs demonstrate the efficacy of the history model in improving the running times of programs on precisely those inputs for which such times are longer. They also demonstrate, for the majority of the cases we investigated, that running times tend to become better by a non-negligible margin than those obtained by O3 or PH optimization.

As becomes apparent from Sections 2 through 4, maintaining a Bayesian-network history model for a program as it is run on a sequence of inputs depends on strategy and parameter choices that are not necessarily unique. This is true of our choice of a geometric average to combine two Bayesian networks, and also of the functional form of (15) to select weights for the geometric average. It is similarly true of our choice method for solving the history model (stochastic simulation coupled with simulated annealing) and of the parameters involved.

But however arbitrary some of these choices are, running an instrumented version of the program for trace recording and solving the history model are costly procedures, so one naturally wonders about the practicality of the overall approach in a real-world context. Our vision here is that the strategy summarized in Steps 1 and 2 of Section 3 is not to be applied to the sequence of all the inputs that come along for execution by program $P$, but rather on a subsequence of that sequence, for example as follows.

When input $I_1$ arrives, two instances of $P$ are started on it. The first instance is not instrumented and returns the result of the execution as soon as it becomes available. The second instance, in turn, is instrumented and yields an execution model to be incorporated into the history model for $P$. New inputs that appear in the meantime only cause one instance of $P$ to be started (the one that is not instrumented). Once the history model for $P$ has been updated and solved, and a corresponding reordered code has been obtained, a new input for $P$ may then once again trigger two executions of $P$, but now employing the newly reordered code. In this vision, a background system can be dedicated to maintaining history models and from time to time releasing versions of crucial programs that are tuned to the types of demand they have encountered.

## Acknowledgments

The authors acknowledge partial support from CNPq, CAPES, and a FAPERJ BBP grant. They also thank S. Debray for providing access to PLTO and for promptly making new updates available.